\documentclass[preprint,12pt]{elsarticle}

\usepackage{color}

\journal{Journal of Alloys and Compounds}

\begin{document}

\begin{frontmatter}


\title{Strain effects on the electronic structure of the FeSe$_{0.5}$Te$_{0.5}$ superconductor}

\author[INT]{M. J. Winiarski}
\author[INT]{M. Samsel-Czeka\l a\corref{cor1}}
\author[IF]{A. Ciechan}
\cortext[cor1]{Corresponding author}
\address[INT]{Institute of Low Temperature and Structure Research, Polish Academy of Sciences, Ok\'olna 2, 50-422 Wroc\l aw, Poland}
\address[IF]{Institute of Physics, Polish Academy of Sciences, al. Lotnik\'ow 32/46, 02-668 Warsaw, Poland}

\begin{abstract}
The electronic structure of the strained FeSe$_{0.5}$Te$_{0.5}$ superconductor has been investigated from first principles. Our calculation results indicate that the influence of hydrostatic, biaxial or uniaxial compressive stress on the density of states at the Fermi level is insignificant. The overall shape of the Fermi-surface (FS) nesting function for FeSe$_{0.5}$Te$_{0.5}$ at ambient pressure resembles that of its parent compound, FeSe, but under the {\it ab}-plane compressive strain. In these two systems, changes of their FSs under various stress conditions are qualitatively almost the same. However, in FeSe$_{0.5}$Te$_{0.5}$ the intensity of the perfect $\mathbf{Q}=(0.5,0.5)\times(2\pi/a)$ nesting vector is more diminished. These findings are in good agreement with former experimental data and support the idea of spin-fluctuation mediated superconductivity in iron chalcogenides.
\end{abstract}


\end{frontmatter}

\section{Introduction}

Iron chalcogenide superconductors are extensively investigated because of their simple crystal structure which enables promising applications. The superconducting transition temperature $T_c=$ 8 K in the pure FeSe \cite{Hsu} but it is risen up to 15 K in the FeSe$_{1-x}$Te$_x$ solid solutions for $x=0.5$ \cite{Yeh, Fang, Mizuguchi}. Then, $T_c$ reaches more than 30 K for the $A_x$Fe$_2$Se$_2$ ternaries with the alkali metals atoms {\it A} (= K, Rb, Cs) located between the Fe-Se layers \cite{Guo, Fang-AFeSe, Luo, Ying}. Furthermore, the maximum $T_c$ of 37 K has been reported for the pure FeSe under as high hydrostatic pressure as 9 GPa \cite{Mizuguchi-FeSe, Margadonna, Millican, Kumar, Okabe}. In FeSe$_{0.5}$Te$_{0.5}$, the $T_c$ reaches its maximum value of 26 K at lower pressure (2 GPa) \cite{Horigane}. An enhancement of $T_c$, among such systems containing tellurium, is achieved in Mn-doped FeSe$_{0.5}$Te$_{0.5}$ crystals \cite{Gunther}. In turn, non-hydrostatic strains also modify $T_c$ of the 11-type 
superconductors as it was revealed in lattice mismatched epitaxial films where tensile strain suppresses superconductivity (SC) \cite{Nie, Huang}. It was an opposite effect to that of the compressive  biaxial ({\it ab}-plane) \cite{Huang, Bellingeri, Bellingeri2} or uniaxial ({\it c}-axis) \cite{Si} strain, causing the increase of the $T_c$'s. 

The strong influence of both chemical and external pressure on superconducting properties of iron chalcogenides can be partly explained by changes of the structural properties. The optimal conditions for SC are closely related to the imperfect tetrahedral coordination of an Fe atom, tuned by a chalcogenide anion height (free $Z_{Se/Te}$ position) \cite{Okabe, Huang, Bellingeri2}.

The electronic structure investigations of superconducting iron chalcogenides, both theoretical \cite{Subedi, Chen, Ciechan, Singh, Chadov} and experimental \cite{Kumar, Chen, Tamai, Nakayama, Miao}, have shown that the multi-gap nature of their SC may be connected with interband interactions between the holelike $\beta$ and electronlike $\delta$ Fermi surface (FS) sheets. In particular, SC can be mediated by antiferromagnetic spin fluctuations (SF), observed experimentally by e.g. NMR \cite{Imai, Shimizu}, which are driven by the imperfect nesting with the $\mathbf{q}\sim(\pi,\pi)$ vector, spanning the above FS sheets in iron chalcogenides \cite{Subedi, Ciechan, Singh, EPL}.

On the one hand, magnetic fluctuations, observed in NMR experiments  \cite{Shimizu}, are stronger in the pure FeSe compared with those in FeSe$_{0.5}$Te$_{0.5}$ while the $T_c$ of the latter compound is much higher. On the other hand, for both compounds, these fluctuations are developing with the external pressure and, at the same time,   their $T_c$'s increase. Since the nesting properties that can determine these phenomena are very subtle, their modifications can be described, both qualitatively and quantitatively, by theoretical studies. Our previous results for the pure FeSe \cite{EPL} showed precise nesting function changes under different pressure and strain conditions being connected with either suppression or enhancement of the SC in this parent compound for all iron chalcogenide superconductors.

In this work, we focus on studying changes of both the crystal parameters and electronic structure of FeSe$_{0.5}$Te$_{0.5}$ with setting various kinds of strain in the unit cell (u.c.). In particular, densities of states (DOS) and FS details are examined. By the use of an estimated nesting function, the FS features of FeSe$_{0.5}$Te$_{0.5}$ are compared with those of the pure FeSe. The relation between the FS-nesting modifications and SF-mediated SC properties in iron chalcogenides is discussed based on available experimental data. We show both a qualitative and quantitative description of the $\mathbf{q}\sim(\pi,\pi)$ nesting vector intensities that implicitly explain the dependence of $T_c$'s in the considered 11-type systems under chemical pressure and/or strains. 

\section{Computational details}

Band structure calculations for FeSe$_{0.5}$Te$_{0.5}$ have been carried out in the framework of the density functional theory (DFT). A full optimization of the atomic positions and geometry of the tetragonal u.c. of the $P4mm$-type (No. 99) under various stress conditions was performed with the Abinit package \cite{Abinit, PAW}, using Projector Augmented Wave (PAW) pseudopotentials, generated with Atompaw software \cite{Atompaw}. The local density approximation (LDA) \cite{JP}, employed here, seems to be adequate for electronic structure calculations for the 11-type systems, as discussed earlier \cite{EPL}. The 3s3p3d;4s4p states for both Fe and Se atoms as well as the 4s4p4d;5s5p states for the Te atom were selected as a valence-band basis. The following three types of compressive strains were considered: hydrostatic pressure, biaxial in the {\it ab}-plane as well as uniaxial in the {\it c}-axis direction compressive stress. Based on these results, the full potential local-orbital (FPLO) band structure 
code \cite{FPLO} was used in the scalar-relativistic mode to compute the DOSs and Fermi surfaces. Since the FS nesting features  of the 11-type compounds are tiny, very dense {\bf k}-point meshes in the Brillouin zone (BZ) had to be used, {\it i.e.} 64$\times$64$\times$64 and 256$\times$256$\times$256 for the self-consistent field (SCF) cycle and FS maps, respectively.

Finally, a nesting function was determined numerically by the formula: 

\begin{equation}
f_{nest}(\mathbf{q})=\Sigma_{\mathbf{k},n,n'}
\frac{[1-F_{n}^{\beta}({\mathbf{k}})]F_{n'}^{\delta}(\mathbf{k+q})}
{|E_{n}^{\beta}({\mathbf{k}})-E_{n'}^{\delta}(\mathbf{k+q})|},
\end{equation}

where $F_{n}^{\beta}$ and $F_{n'}^{\delta}$ are the Fermi-Dirac functions of states $n$ and $n'$ in bands $\beta$ and $\delta$, ($F=$ 0 or 1 for holes or electrons), respectively. $E_{n}^{\beta}$ and $E_{n'}^{\delta}$ are energy eigenvalues of these bands. The studied $f_{nest}({\mathbf{q||Q}})$, were $\mathbf{Q}=(0.5,0.5)\times(2\pi /a)$ is the ideal nesting vector, represents a frequency of an occurrence of a given vector $\mathbf{q}\sim(\pi,\pi)$ (having its length close or equal to that of $\mathbf{Q}$) in the {\bf k}-space, spanning the FS sheets originating from the $\beta$ and $\delta$ bands.

\section{Results and discussion}

The structural parameters calculated for the unstrained FeSe$_{0.5}$Te$_{0.5}$, compared with other theoretical and experimental results, available in the literature, are collected in table 1. As this table indicates, structural parameters, presented here, differ somewhat from results of our former LDA study \cite{Ciechan}. This effect is related to the size of a valence basis of PAW pseudopotentials. In this work, additional 3s3p states for Fe/Se atoms and 4s4p states for a Te atom were included into calculations, leading to slightly higher values of {\it a} and significantly enhanced values of {\it c} parameters. In general, LDA lattice parameters have usually lower values than corresponding experimental data, thus an underestimated by 3.1\% value of {\it a} is a standard result. Meanwhile, almost an ideal LDA value, obtained for the {\it c} parameter can be explained by the layered character of the iron chalcogenides structure, where a metallic bond is formed mainly by the Fe d-electrons, whereas the 
interlayer, covalent bond is connected with p-electrons coming from Se and Te atoms. Here, more complete PAW pseudopotentials describe better only the covalent bond in the {\it c}-axis direction without changes made in the {\it ab}-plane with respect to the previous LDA results. Hence, theoretical optimizations of forces, considered here, may be affected by systematic, anisotropic discrepancies between a description of the {\it ab}-plane and interplane interactions.

Electronic structure changes (particularly the Fermi surface nesting), investigated here, are closely related to the Se/Te anion heights and Fe-Se/Fe-Te bond lengths. Calculated here free atomic positions, $Z_{Se}$ and $Z_{Te}$, for the unstrained FeSe$_{0.5}$Te$_{0.5}$ deviate from experimental ones by -7\% and 2.8\%, respectively. The least satisfying result, obtained for $Z_{Se}$, is also reflected in significant underestimation of the Fe-Se bond length, despite that the Fe-Te bond length, determined here, is perfect. All these effects lead to a description of structural and electronic properites that are related to stronger compressively strained (in {\it ab}-plane) systems than the real FeSe$_{0.5}$Te$_{0.5}$. However, these disadvantages of the LDA method performance in such an anisotropic system are meaningless to formulate main conclusions of this study that have rather a qualitative character.

For the FeSe$_{0.5}$Te$_{0.5}$ superconductor, stress dependencies of its optimized lattice parameters {\it a} and {\it c}, as well as free atomic $Z_{Se/Te}$ positions are displayed in figures 1 and 2. It appears that the effect of particular strains on the above parameters is very similar to that in FeSe  \cite{EPL}. Namely, the lattice parameter {\it a} (associated with the Fe atoms network) only slightly changes while the {\it c/a} ratio decreases. As is seen in figures 1 and 2, the most significant variations of the lattice parameters and free $Z_{Se/Te}$ positions are caused by the {\it c}-axis compressive stress.

It is known that $T_c$ of iron chalcogenides increases up to its maximum value for pressure of 2 GPa and decreases monotonically for higher pressure \cite{Horigane}. Interestingly, the changes of the anion height caused by any kind of stress are similarly nonlinear, as seen in fig. 2. At the same time, the relation between the $T_c$ and $Z_{Se/Te}$ (and also Fe-Se and Fe-Te bond lengths) is clearly linear, as was reported for FeSe$_{0.5}$Te$_{0.5}$ thin films deposited on e.g., LaAlO$_3$ or SrTiO$_3$ \cite{Bellingeri, Bellingeri2}. Hydrostatic pressure of 2 GPa, investigated here, is related just to the maximum of bulk $T_c$ \cite{Horigane}, whereas other values of stress were chosen arbitrarily. However, some combination of compressive biaxial and uniaxial stress conditions, considered in this study, can lead to structural changes analogous to those present in real thin films. The results of this work may be also a good approximation of electronic structure of strained FeSe$_{0.5}$Te$_{0.5}$ systems.
 
As is visible in figure 3, the DOSs at the Fermi level, N(E$_F$), are stable under various stress conditions. Therefore, changes of $T_c$ in FeSe$_{0.5}$Te$_{0.5}$ cannot be associated with modifications of N(E$_F$).  Meanwhile, in the spin-fluctuation mediated SC scenario, the FS nesting $\mathbf{q}\sim(\pi,\pi)$, schematically presented in figure 4, is crucial for the superconducting pairing.

In superconducting FeSe$_{0.5}$Te$_{0.5}$, chemical pressure introduced by the Te-atom substitution into FeSe, leads to the overall shape of the nesting function being similar to that of the pure FeSe under {\it ab}-plane compressive strain of 3 GPa, as illustrated in fig. 5 a) and e) (see also fig. 7 c) of Ref. \cite{EPL}). On the one hand, this effect is related to the structural anisotropy of the 11-type systems, where the {\it ab}-plane and interplane interactions are different. On the other hand, the problem of structure optimization, mentioned above, influences these results. Interestingly, as is seen in figure 5 parts b)-d), overall $f_{nest}$ of FeSe$_{0.5}$Te$_{0.5}$ under all considered here kinds of compressive stress of magnitude 1 or 2 GPa is further broadened and the intensity of peaks around  $\mathbf{Q}$ are somewhat diminished and shifted towards smaller values of  $|\mathbf{q}|$. These effects resemble qualitatively those reported for FeSe \cite{EPL} but, after primary modifications caused 
by the chemical pressure, here they are not so rapid. Furthermore, the stress induced changes of T$_c$ in FeSe$_{0.5}$Te$_{0.5}$ are also less rapid, compared with pure FeSe \cite{Horigane}.

Previous theoretical investigations suggested the ordered spin density wave (SDW) state, being more energetically favourable in FeSe$_{0.5}$Te$_{0.5}$ than in FeSe \cite{Subedi}. On the contrary, the intensity of the perfect nesting $\mathbf{Q}$ vector in unstrained FeSe$_{0.5}$Te$_{0.5}$ (fig. 5 a)) is diminished, which suggests that the ordered SDW state, determined by the $\mathbf{Q}$ vector is significantly suppressed. At the same time, vectors with lengths close to $\mathbf{Q}$ seem to be responsible for SF of the antiferromagnetic character occurring now in a broader energy range. Our description of the electronic structure for 11-type superconductors implicates that two effects may influence the SC phenomenon in these systems. First, the suppression of $f_{nest}(\mathbf{Q})$, which is exhibited in FeSe and FeSe$_{0.5}$Te$_{0.5}$ under various compressive strain conditions, leads to disappearing of the ordered SDW state. Second, the simultaneous increase of the imperfect $\mathbf{q}\sim(\pi,\pi)$, but 
close to that of $\mathbf{Q}$, vectors intensities should considerably enhance SF of an antiferromagnetic character, being responsible for SC. Furthermore, the more uniform shape of total $f_{nest}$, obtained for stressed iron chalcogenides, makes the SF energy scale wider, which may also influence the $T_c$.

These findings are in good accord with former experimental reports on antiferromagnetic SF in FeSe and FeSe$_{0.5}$Te$_{0.5}$ \cite{Shimizu}. In turn, some compressive strains enhance the superconductivity in FeSe$_{0.5}$Te$_{0.5}$ thin films \cite{Bellingeri, Bellingeri2}. Nevertheless, exact relations between the ordered SDW state, the intensity of SF, and $T_c$'s of 11-type systems remain unclear. The strength of superconducting pairing in this 11-family of compounds seems to be system dependent and indirectly related to the particular FS nesting intensity, which can explain some disproportions of $T_c$ between FeSe and FeSe$_{0.5}$Te$_{0.5}$. 

\section{Conclusions}

The electronic structure of superconducting FeSe$_{0.5}$Te$_{0.5}$ has been studied under various stress conditions. The Fermi surface nesting character of unstrained FeSe$_{0.5}$Te$_{0.5}$ was found to be similar to that of FeSe under {\it ab}-plane stress of 3 GPa. In general, various strain effects on electronic structure of these 11-type systems are qualitatively the same. The intensity of the perfect nesting with $\mathbf{Q}$ in FeSe$_{0.5}$Te$_{0.5}$ is diminished, suggesting suppression of the SDW ordering, while the superconducting transition temperature is considerably enhanced. These findings support the assumption of spin-fluctuation mediated SC in iron chalcogenides, but also reveal the system-dependent strength of SF in mediating the superconducting pairing, which requires further studies.

\section*{Acknoledgements}
This work has been supported by the EC through the FunDMS Advanced Grant of the European Research Council (FP7 âIdeasâ) as well as by the National Center for Science in Poland (Grant No. N N202 239540 and Grant No. DEC-2011/01/B/ST3/02374). The calculations were partially performed on the ICM supercomputers of Warsaw University (Grant No. G46-13) and in Wroc\l aw Centre for Networking and Supercomputing (Project No. 158).

\begin{table}
\caption{Calculated lattice parameters {\it a} and {\it c}, free atomic positions, $Z_{Se/Te}$ and corresponding bond lengths (d$_{Fe-Se}$/d$_{Fe-Te}$), in unstrained u.c. of FeSe$_{0.5}$Te$_{0.5}$.}
\label{table1}
\begin{tabular}{lllllll}
reference &  {\it a} (nm) & {\it c} (nm) & $Z_{Se}$ & $Z_{Te}$ & d$_{Fe-Se}$ (nm) & d$_{Fe-Te}$ (nm) \\ \hline
this work (LDA) & 0.3681 & 0.6008 & 0.227 & 0.293 & 0.2291 & 0.2547 \\
theor. ref. \cite{Ciechan}& 0.3655 & 0.5685 & 0.238 & 0.289 & 0.2274 & 0.2457 \\
exp. ref \cite{Louca}& 0.3800 & 0.5954 & 0.244 & 0.285 & 0.2392 & 0.2547\\
$\Delta_{LDA-exp}$ (\%) & -3.13 & 0.91 & -6.97 & 2.81 & -4.22 & -0.02 \\
\end{tabular}
\end{table}

\begin{figure}
\includegraphics[scale=0.8]{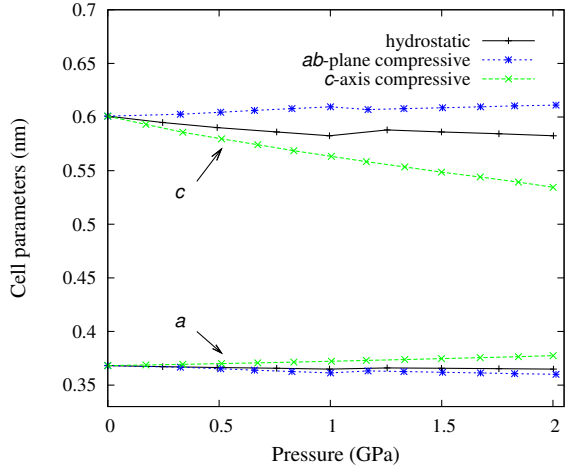}
\caption{Variations of lattice parameters {\it a} and {\it c} of tetragonal FeSe$_{0.5}$Te$_{0.5}$ u.c. with different strain conditions.}
\label{Fig1}
\end{figure}

\begin{figure}
\includegraphics[scale=0.8]{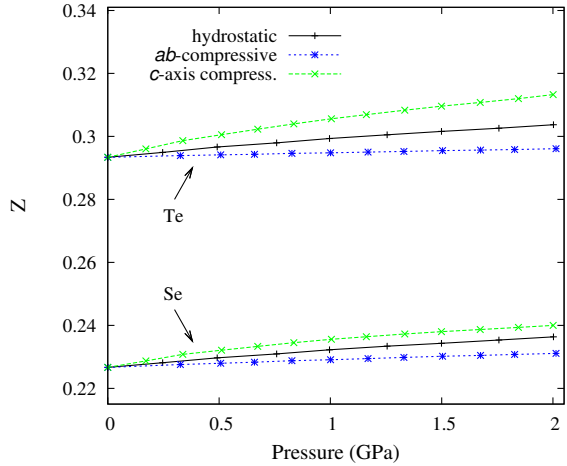}
\caption{The same as in figure 1 but for the free atomic $Z_{Te}$ and $Z_{Se}$ positions.}
\label{Fig2}
\end{figure}

\begin{figure}
\includegraphics[scale=0.8]{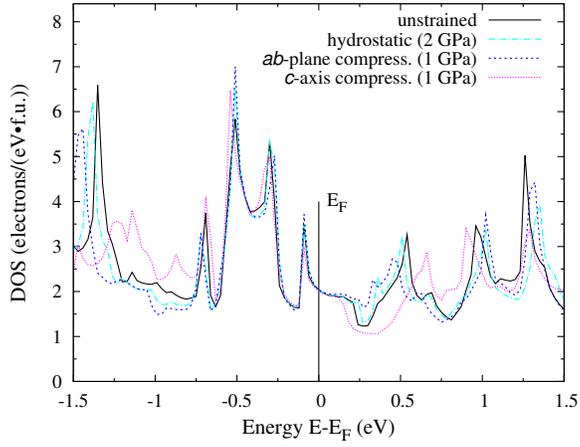}
\caption{Total DOS's plots of FeSe$_{0.5}$Te$_{0.5}$ under various strain conditions.}
\label{Fig3}
\end{figure}

\begin{figure}
\includegraphics[scale=0.8]{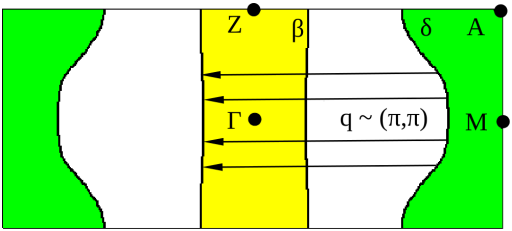}
\caption{Nesting vectors, $\mathbf{q}\sim(\pi,\pi)$, spanning the $\beta$ and $\delta$ FS sheets, marked on $\Gamma$-$M$-$A$-$Z$ FS section of unstrained FeSe$_{0.5}$Te$_{0.5}$}
\label{Fig4}
\end{figure}

\begin{figure}
\includegraphics[scale=0.7]{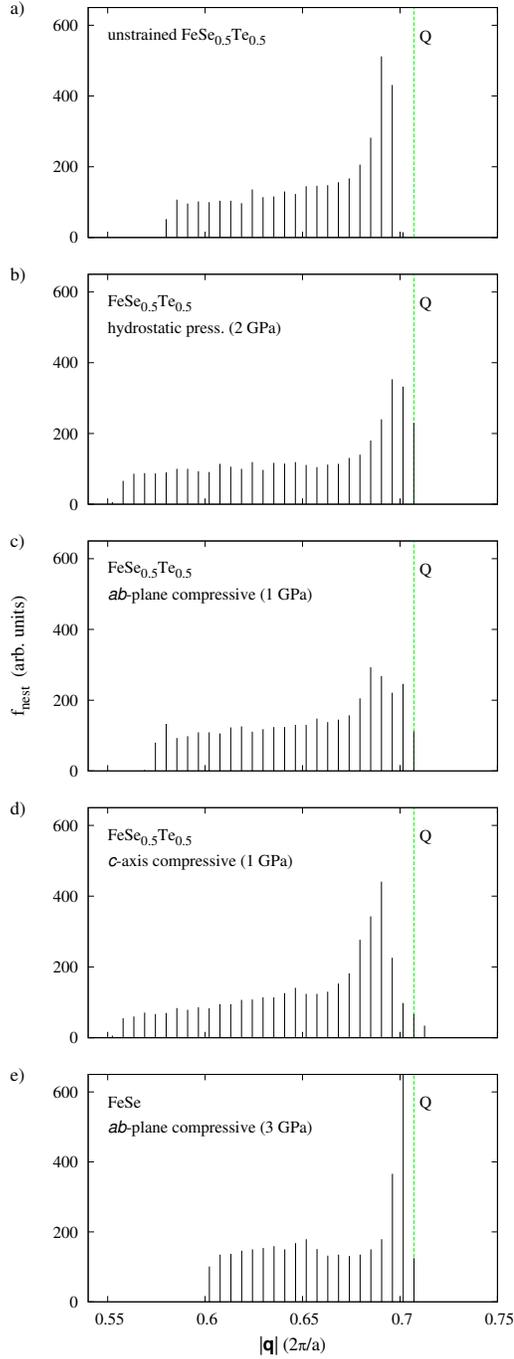}
\caption{Histograms representing the nesting function, $f_{nest}$ vs. lengths of possible spanning vectors $\mathbf{q}$ for the $\beta$ and $\delta$ FS sheets of FeSe$_{0.5}$Te$_{0.5}$  under different kinds of strain, compared with  {\it ab}-plane compressive strained FeSe. The length of the ideal vector $\mathbf{Q}$ (0.7071) is marked by vertical dashed lines.}
\label{Fig5}
\end{figure}

\end{document}